\documentclass[twocolumn, aps,floats,letterpaper,floatfix,groupedaddress]{revtex4-1}
\usepackage{amsmath}
\usepackage{graphicx}
\usepackage{amsfonts}
\usepackage{amssymb}
\usepackage{bm}
\usepackage{dsfont}
\usepackage{epsfig,float,afterpage}
\usepackage[colorlinks,linkcolor=blue,citecolor=blue,urlcolor=blue]{hyperref}
\usepackage[usenames,dvipsnames]{color}

\newcommand{\beq}{\begin{equation}}
\newcommand{\eeq}{\end{equation}}
\newcommand{\bes}{\begin{subequations}}
\newcommand{\ees}{\end{subequations}}
\newcommand{\bea}{\begin{eqnarray}}
\newcommand{\eea}{\end{eqnarray}}
\newcommand{\ba}{\begin{array}}
\newcommand{\ea}{\end{array}}
\newcommand{\beqn}{\begin{eqnarray*}}
\newcommand{\eeqn}{\end{eqnarray*}}

\newcommand{\f}[2]{\frac{#1}{#2}}

\newcommand{\la}{\langle}
\newcommand{\ra}{\rangle}

\newcommand{\dg}{\dagger}

\newcommand{\Hh}{\hat{H}}
\newcommand{\MM}{\mathcal{M}}
\newcommand{\MN}{\mathcal{N}}
\newcommand{\tr}{\text{tr}}

\def\one{1\hskip -1mm{\rm l}}

\def\nn{\nonumber}

\begin{document}
\title{Spectral form factor in a minimal bosonic model of many-body quantum chaos}
\author{Dibyendu Roy$^1$, Divij Mishra$^2$ and Toma$\check{\rm z}$ Prosen$^3$}\affiliation{$^1$Raman Research Institute, Bangalore 560080, India}
\affiliation{$^2$Indian Institute of Science, Bangalore 560012, India}
\affiliation{$^3$Physics Department, Faculty of Mathematics and Physics, University of Ljubljana, Jadranska 19, SI-1000 Ljubljana, Slovenia} 

\begin{abstract}
We study spectral form factor in periodically-kicked bosonic chains. We consider a family of models where a Hamiltonian with the terms diagonal in the Fock space basis, including random chemical potentials and pair-wise interactions, is kicked periodically by another Hamiltonian with nearest-neighbor hopping and pairing terms. We show that for intermediate-range interactions, random phase approximation can be used to rewrite the spectral form factor in terms of a bi-stochastic many-body process generated by an effective bosonic Hamiltonian.
In the particle-number conserving case, i.e., when pairing terms are absent, the effective Hamiltonian has a non-abelian $SU(1,1)$ symmetry, resulting in universal quadratic scaling of the Thouless time with the system size, irrespective of the particle number. This is a consequence of degenerate symmetry multiplets of the subleading eigenvalue of the effective Hamiltonian and is broken by the pairing terms. In the latter case, we numerically find a nontrivial systematic system-size dependence of the Thouless time, in contrast to a related recent study for kicked fermionic chains.
\end{abstract}

\maketitle
\section{Introduction}
 Understanding chaos in many-body quantum systems with or without classical limit has received significant renewed interest in recent years \cite{Prosen2007JPA, Akila2016JPA, Yoshida2016JHEP, MaldacenaPRD2016, MaldacenaJHEP2016, Bohrdt_2017,CarlosPRL2019, KosPRX2018, BertiniPRL2018, Bertini2021CMP, ChanPRL2018, ChanPRX2018, Gharibyan2018, FriedmanPRL2019, RoyPRE2020, LiPRL2021, Moudgalya21, KosPRL2021,  Garratt2021PRX, Garratt2021PRL, Liao21}. The study of quantum chaos and its connection to random matrix theory (RMT)~\cite{Haake2001, Fyodorov2011} is essential in the description of ergodicity and thermalization in closed quantum systems~\cite{AnatoliReview, Rigol2008Nature, Bertini2019PRL, Arul2021PRR, Lamacraft2021PRL, Bertini2021PRX}. While many new concepts, such as out-of-time-ordered correlators \cite{MaldacenaPRD2016,MaldacenaJHEP2016,Bohrdt_2017,CarlosPRL2019}, the growth of entanglement entropy and operator spreading \cite{NahumPRX2017, Keyserlingk2018, Nahum2018PRX, Khemani2018PRX, BertiniPRX2019}, have been explored for the identification of chaotic quantum dynamics, the statistical description of energy or quasienergy spectra of complex quantum systems remains one of the main signatures of quantum chaos \cite{Haake2001, Sieber2001, Sieber2002}. The spectral form factor (SFF) $K(t)$, a measure of spectral fluctuations, has been investigated analytically in a series of recent studies \cite{KosPRX2018, BertiniPRL2018, Bertini2021CMP, ChanPRL2018, ChanPRX2018} to formulate ideas of quantum chaos in strongly interacting, nonintegrable systems where local degrees of freedom have no classical limit. These studies have identified dynamical mechanisms for the emergence of RMT description of the spectral properties of many-body systems by going beyond the semiclassical periodic-orbit approaches \cite{Berry1977, Berry1985, Sieber2001,Sieber2002,MullerPRL2004,MullerPRE2005}. 
 
 The models considered in these early papers  \cite{KosPRX2018, BertiniPRL2018, ChanPRL2018, ChanPRX2018} do not have any conserved quantity due to the unitary symmetry of the system whose role in many-body quantum chaos was later explored in Ref.~\cite{FriedmanPRL2019} in a Floquet circuit model with a large local Hilbert space and without time-reversal invariance. The role of $U(1)$ symmetry (particle-number conservation) was further investigated by two of us in a one-dimensional (1D) lattice of interacting spinless fermions with a time-periodic kicking in the nearest-neighbor coupling (hopping and/or pairing) \cite{RoyPRE2020}. The fermionic model has a finite local Hilbert space and possesses a time-reversal symmetry. In Ref.~\cite{RoyPRE2020}, we have suggested a new dynamical chaos mechanism that maps $K(t)$ to an average recurrence probability of a classical Markov chain with transition probabilities given as square-moduli of hopping (and pairing) amplitudes. We show here that such a mechanism is a general one for Floquet lattice models with long- or intermediate-range pairwise interactions and random diagonal terms allowing random phase approximation (RPA). To demonstrate that, we study a bosonic version of the model explored in \cite{RoyPRE2020}, which complements the existing studies of many-body quantum chaos with spins and fermions.

 The main difference of the bosonic model from the fermionic one arises from the unrestricted number of spinless bosons (constrained only by the total number of bosons in the lattice) at any site. The infinite-dimensional local Hilbert space of bosons poses a challenge to numerically explore $K(t)$ and the Thouless time in the bosonic chains, especially in the absence of $U(1)$ symmetry. Further, one might expect different scaling of Thouless time with system sizes for bosons in contrast to fermions due to the differences in their statistics. The Thouless timescale beyond which $K(t)$ has the universal form of RMT scales asymptotically with the system size $L$ in the fermionic kicked lattice as ${\cal O}(L^2)$, or ${\cal O}(L^0)$, in the presence, or absence of particle-number conservation, respectively. Surprisingly, we find the same scaling of Thouless time with system sizes for the particle number conserving bosonic model. We argue this similarity between fermionic and bosonic models is due to the universal non-abelian symmetry of the underlying Markov matrices whose subleading eigenvalues determine the scaling of Thouless time. We identify the corresponding symmetry groups as compact $SU(2)$ and non-compact $SU(1,1)$ for the fermionic and bosonic models. In the presence of these symmetries, the subleading eigenvalues (mainly, the second largest eigenvalue) of Markov matrices and their system-size dependence are independent of the number of fermions or bosons in the entire chain as they are descendants of the single-particle eigenvalues through the degenerate symmetry multiplets. Since the Markov matrix is identical for the single-particle fermionic or bosonic system, we find the same $L$-scaling of Thouless time for fermions and bosons in the presence of $U(1)$ symmetry. The scaling of Thouless time in the bosonic model in the absence of $U(1)$ symmetry suggests a systematic system-size dependence, which is different from the fermionic model.
 
\section{Model and spectral form factor} 
Following the study on the fermionic chain~\cite{RoyPRE2020}, we here investigate a 1D lattice of interacting spinless bosons with a time-periodic kicking in the nearest-neighbor coupling (hopping). The full Hamiltonian reads as (we set $\hbar=1)$
\bea
\hat{H}(t)&=&\hat{H}_0+\hat{H}_1\sum_{m \in \mathbb{Z}}\delta(t-m),\label{ham1} \\
\hat{H}_0&=& \sum_{i=1}^L\epsilon_i\hat{n}_i + \sum_{i<j} U_{ij}\hat{n}_i\hat{n}_j, \\
\hat{H}_1&=&\sum_{i=1}^L(-J\hat{a}_i^{\dg}\hat{a}_{i+1}+\Delta\hat{a}_{i}^{\dg}\hat{a}^{\dg}_{i+1}+{\rm H.c.}),
\eea
where the time is measured in units of pulse period (cycle). Here, $\hat{n}_i=\hat{a}_{i}^{\dg}\hat{a}_{i}$ is the number operator where $\hat{a}_{i}^{\dg}$ is a creation operator of a boson at site $i$.  We use periodic boundary conditions (PBC) in real space, i.e., $\hat{a}_i\equiv\hat{a}_{i+L}$.  
The long-range interaction between bosons at sites $i$ and $j$ is given by $U_{ij}=U_0/d(i,j)^\alpha$, $d(i,j) = {\rm min}(|i-j|,|i-j+L|,|i-j-L|)$,
 with an exponent in the interval $1<\alpha<2$, and the random onsite energies $\epsilon_i$ described as Gaussian {\em i.i.d.} variables of zero mean and standard deviation $\Delta \epsilon$. We consider the driving Hamiltonian $\hat{H}_1$ with or without a $U(1)$ symmetry which corresponds respectively to conservation or violation of a total boson number $\hat{N}=\sum_{i=1}^L\hat{n}_i$. The strength of hopping and the amplitude of pairing (creation or annihilation of a boson pair) are respectively $J$ and $\Delta$. The absence or presence of pairing $\Delta$ generates $U(1)$ symmetric or symmetry-broken kicking. The bosonic model~\eqref{ham1} can be realized with photons in optical systems \cite{EckardtRMP2016, RoyRMP2017}, where the pairing term can be mediated through two-photon processes (e.g., parametric amplification or down-conversion) in nonlinear optical medium with second-order susceptibility \cite{scully1997quantum}. 

We define the SFF as
\bea
K(t)=\langle (\tr \hat{U}^t)(\tr \hat{U}^{-t}) \rangle - \MN^2 \delta_{t,0}, \label{SFF}
\eea
where $\MN$ is the dimension of the Hilbert space of the bosonic chain, and $\langle \dots \rangle$ denotes an average over the quench disorder $\{ \epsilon_i\}$. The SFF $K(t)$ in Eq.~\ref{SFF} without the additional averaging $\langle \dots \rangle$ over disorder (an ensemble of similar systems) is not a self-averaging quantity. In the absence of disorder in the model, such disorder averaging can be replaced by an appropriate additional averaging over local windows of time (moving time average) to make $K(t)$ self-averaging \cite{KosPRX2018}. The one-cycle time-evolution operator $\hat{U}$ can be expressed as
\bea
\hat{U} = \hat{V}\hat{W}, \quad
	\hat{W} = e^{-i\Hh_0} \text{ and } \hat{V} = e^{-i\Hh_1}.
        \eea
 To evaluate $K(t)$, we choose a basis of Fock states $|\underline{n}\ra \equiv |n_1,n_2,\dots,n_L\ra$, where the occupation number of spinless boson at the lattice site $j$ is given $n_j\in \{0,1,\dots,N\}$ with a constraint $N \equiv \la\underline{n}|\hat{N}|\underline{n}\ra=\sum_{j=1}^Ln_j$.
 For $\Delta=0$, since $[\hat{U},\hat{N}]=0$, we consider ${\cal N}=\frac{(N+L-1)!}{N!(L-1)!}$ dimensional Hilbert subspace with fixed total number $N$ of bosons.
On the other hand, when $\Delta \ne 0$, the Hilbert space is formally infinite dimensional for any $L$. 
In order to obtain meaningful approximate numerical results, we truncate the Hilbert space by considering all even or odd $N$ bases with varying $N$ up to some cutoff $N_{\rm max}$. The truncated Hilbert space dimension for even bases with an even $N_{\rm max}$ is then ${\cal N}=\sum_{N=0,2,4,\dots}^{N_{\rm max}}\frac{(N+L-1)!}{N!(L-1)!}$, where (in)dependence of the results on   $N_{\rm max}$ needs to be carefully investigated.

\begin{figure}
\includegraphics[width=\linewidth]{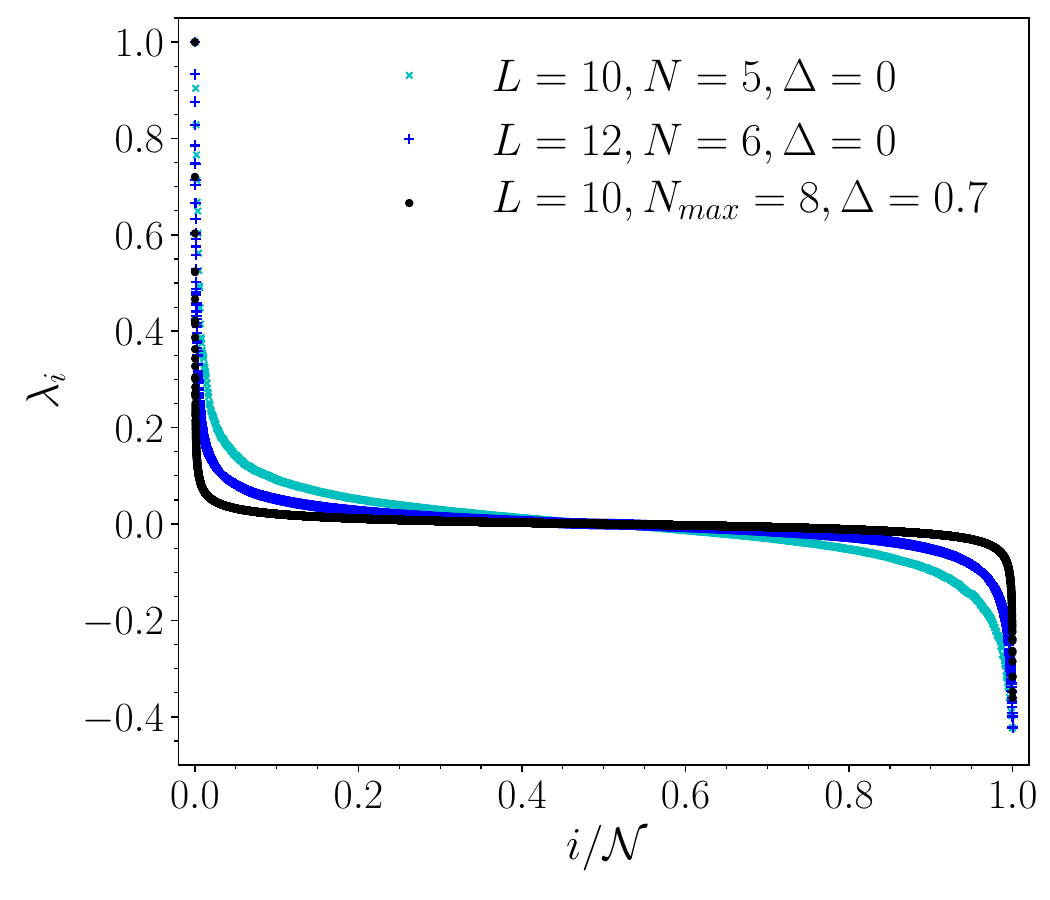}
\caption{Eigenvalues $\lambda_i$ of the doubly stochastic matrix $\mathcal{M}$ for different system sizes $L$ and number of bosons $N$ or $N_{\text{\rm max}}$ in the presence $(\Delta=0)$ and absence $(\Delta \ne 0)$ of $U(1)$ symmetry in the kicked interacting bosonic chain with $J=1$.}
\label{EvBoson}
\end{figure}

In either case of $\Delta=0$ or $\Delta\ne 0$, the Fock basis states $|\underline{n}\ra$ are eigenstates of $\Hh_0$ and $\hat{W}$:
\bea
\hat{W}|\underline{n}\ra=e^{-i\theta_{\underline{n}}}|\underline{n}\ra,~\theta_{\underline{n}}=\sum_{i=1}^L\epsilon_in_i + \sum_{i<j} U_{ij}n_in_j,
\eea
where the phases $\theta_{\underline{n}}$ for different many-particle basis states $|\underline{n}\rangle$ (modulo $2\pi$) are approximated as independent uniformly distributed random numbers. This allows us to use RPA to perform the disorder averaging over different realizations. 
We further make the asymptotic approximation via dihedral subgroup $D_{2t}$ of permutations between two replicas~\cite{KosPRX2018} to achieve the following simple form  of the SFF for bosons, analogous to fermions~\cite{RoyPRE2020}:
\bea
K(t)=2t\: \tr \MM^t, \label{SP}
\eea
where $\MM$ is a $\mathcal{N} \times \mathcal{N}$  square matrix whose elements are
\bea
\MM_{\underline{n},\underline{n}'}=|\la \underline{n}|\hat{V}|\underline{n}'\ra|^2. \label{Mmat}
\eea
The elements of $\MM$ are non-negative real numbers, and
\bea
\sum_{\underline{n}'}\MM_{\underline{n},\underline{n}'}=\sum_{\underline{n}'}\la \underline{n}|\hat{V}|\underline{n}'\ra \la \underline{n}'|\hat{V}^{\dg}|\underline{n}\ra=\la \underline{n}|\hat{V}\hat{V}^{\dg}|\underline{n}\ra=1, \nn
\eea
where we use the unitarity of $\hat{V}$ at the end. Thus, $\MM$ is a doubly stochastic (Markov) matrix as the sums of its non-negative elements along any rows and columns are equal to $1$. It is known that the largest eigenvalue of a doubly stochastic matrix is 1, while the rest of the eigenvalues are real as $\MM$ is also real and symmetric. We write these eigenvalues of $\mathcal{M}$ as $1,\lambda_1,\lambda_2,\dots$ with $1\ge |\lambda_j| \ge |\lambda_{j+1}|$. In Fig.~\ref{EvBoson}, we show typical eigenvalues $\lambda_i$ in the presence $(\Delta=0)$ and absence $(\Delta \ne 0)$ of $U(1)$ symmetry, which depict a rapid fall of $\lambda_i$ from its maximum value of one with increasing $i$. We then further approximate $K(t)$ at long times $t$, $1\ll t\ll \mathcal{N}$, by keeping up to the second largest eigenvalue $\lambda_1$ of $\mathcal{M}$ \footnote{We provide justification of such approximation for the universal RMT behavior of $K(t)$ in Appendix~\ref{appA}.}. Thus, we obtain for SFF
\bea
K(t)&\simeq& 2t(1+\lambda_1^t) \nn\\
&\simeq& 2t(1+(1-1/t^*(L))^t)\simeq 2t(1+e^{-t/t^*(L)}), \label{RMT} 
\eea
where we take the scaling of $\lambda_1$ with system size $L$ as $1-1/t^*(L)$ following Ref.~\cite{RoyPRE2020}. Here, $t^*$ is the Thouless time beyond which the SFF has a universal RMT/COE form as $K(t)\simeq 2t$.

For the fermionic chain, the $L$-dependence of $t^*$ was found to be ${\cal O}(L^2)$ or ${\cal O}(L^0)$, in the presence, or absence of $U(1)$ symmetry, respectively \cite{RoyPRE2020}. These system-size scalings were obtained by mapping the matrix $\mathcal{M}$ in the Trotter regime at small $J, \Delta$ to a hermitian ``quantum'' Hamiltonian of the isotropic and anisotropic Heisenberg model with PBC in the presence, or absence of $U(1)$ symmetry, respectively. The eigenenergy spectrum of the isotropic Heisenberg model is gapless, and its first ``excited state'' goes as $1-c_1/L^2$ where $c_1$ is a constant. This explains the quadratic $L$-dependence of $t^*$ in the presence of $U(1)$ symmetry in the fermionic chain when $\Delta=0$. 
The anisotropic Heisenberg model has a finite and system-size independent gap in the energy spectrum between the ground and first excited state. Thus, we have a finite and $L$ independent Thouless time for $\Delta \ne 0$. The above analytical predictions of $L$-scaling of $t^*$ (and the corresponding $\lambda_1$) in the Trotter regime were also numerically verified from the $L$-dependence of the eigenvalues of $\mathcal{M}$ matrix for arbitrary $J,\Delta$ in \cite{RoyPRE2020}.

Moreover, the quadratic $L$-dependence of $\lambda_1$ and therefore that of $t^*$ were observed in Ref.~\cite{RoyPRE2020} for all filling fractions $N/L$ (including single fermion case $N=1$) when $\Delta=0$ in the fermionic chain. We argue that this is due to the $SU(2)$ symmetry of the isotropic Heisenberg model in the Trotter regime, which results in the same eigenvalue of first excited states (single magnon states) of the model in different magnetization sectors. We further find numerically that the matrix $\mathcal{M}$ has $SU(2)$ symmetry for arbitrary $J$ when $\Delta=0$. For this, we numerically construct the following operators satisfying $SU(2)$ algebra, $\sigma^{\alpha}=\sum_j\sigma^{\alpha}_j$ with $\alpha \in \{+,-,0\}$ in the fermionic Fock basis states in which $\mathcal{M}$ is also formed. Here, $\sigma^{0}_j=\sigma^z_j/2, \sigma^{\pm}_j=(\sigma^x_j\pm i\sigma^y_j)/2$ and $\sigma^{x,y,z}$ are Pauli matrices at site $j$. We then explicitly check commutation of all $\sigma^{\alpha}$ with $\mathcal{M}$ of the fermionic model at arbitrary $J$ when $\Delta=0$.


\section{Hamiltonian form of Markov matrix}
\label{map}
The derivation of a Hamiltonian form for the Markov matrix $\mathcal{M}$ in the Trotter regime is a bit challenging for bosons compared to fermions, for which the Jordon-Wigner transformation between spinless fermions and spin-1/2 is useful. Here, we give a general method to find the Hamiltonian form in the Trotter regime, and the method is applicable for fermions, bosons, and spins. We can write $\MM$ in Eqs.~(\ref{SP},\ref{Mmat}) using an element-wise commutative product (also known as the Hadamard product) of $\hat{V}$ with $\hat{V}^*$ in the basis $|\underline{n}\ra$. We here denote such product by $\bullet$: $(A \bullet B)_{\underline{m},\underline{n}}=A_{\underline{m},\underline{n}}B_{\underline{m},\underline{n}}$. Further, we expand $\hat{V}$ in the Trotter regime of small parameters of the Hamiltonian $\Hh_1$ to get a Hamiltonian form of $\MM$:
\bea
\MM&=&e^{-i\Hh_1}\bullet e^{i\Hh_1}\nn\\&=&(\one-i\Hh_1-\f{1}{2}\Hh_1^2+\dots)\bullet(\one+i\Hh_1-\f{1}{2}\Hh_1^2+\dots)\nn\\&=&\one+\Hh_1\bullet \Hh_1-\Hh_1^2 \bullet \one + {\cal O}(\Hh_1^4),\label{Trotter}
\eea
where the term $\Hh_1\bullet \Hh_1$ is an element-wise square of $\Hh_1$ and it has non-zero entries at the same positions as $\Hh_1$ but those entries are squared. The other term $\Hh_1^2 \bullet \one$ represents the diagonal entries of $\Hh_1^2$. Below, we derive quantum Hamiltonian  that can generate the matrix $\MM$ in the Trotter regime for the kicked interacting bosonic lattice. 

We divide the driving Hamiltonian in two parts as $\hat{H}_1=\hat{H}_{J}+\hat{H}_{\Delta}$, where
\bea
\hat{H}_{J}=\sum_{i=1}^L(-J\hat{a}_i^{\dg}\hat{a}_{i+1}+{\rm H.c.}),~ \hat{H}_{\Delta}=\sum_{i=1}^L(\Delta\hat{a}_{i}^{\dg}\hat{a}^{\dg}_{i+1}+{\rm H.c.}).\nn
\eea
Since, $\la \underline{n}'|\hat{H}_{J}|\underline{n}\ra$ and $\la \underline{n}'|\hat{H}_{\Delta}|\underline{n}\ra$ are not simultaneously non-zero for any choice of $|\underline{n}\ra$, $|\underline{n}'\ra$, we have $\hat{H}_{J} \bullet \hat{H}_{\Delta}=\hat{H}_{\Delta}\bullet \hat{H}_{J}= 0$, which can be applied to simplify  
\bea
\hat{H}_1\bullet\hat{H}_1=\hat{H}_{J} \bullet \hat{H}_{J}+\hat{H}_{\Delta} \bullet \hat{H}_{\Delta}. \label{Had1}
\eea
We next explain how we obtain these different terms to get the mapping of $\MM$ for the bosonic chain in the Trotter regime. The non-zero matrix elements of $\la \underline{n}'|\hat{a}_i^{\dg}\hat{a}_{i+1}|\underline{n}\ra$ would be $\sqrt{n'_i}\sqrt{n_{i+1}}$, and the corresponding matrix elements in $\hat{H}_{J} \bullet \hat{H}_{J}$ are $n'_in_{i+1}$, which can be generated by the operator $\sqrt{\hat{n}_i}\hat{a}_i^{\dg}\hat{a}_{i+1}\sqrt{\hat{n}_{i+1}}$ in $\MM$. Thus, we write
\bea
\hat{H}_{J} \bullet \hat{H}_{J}=J^2\sum_{i=1}^L(\sqrt{\hat{n}_i}\hat{a}_i^{\dg}\hat{a}_{i+1}\sqrt{\hat{n}_{i+1}}+\sqrt{\hat{n}_{i+1}}\hat{a}_{i+1}^{\dg}\hat{a}_{i}\sqrt{\hat{n}_{i}}).\nn \\\label{HadJ}
\eea
Similarly, we can find the pairing term in Eq.~\eqref{Had1}:
\bea
\hat{H}_{\Delta} \bullet \hat{H}_{\Delta}&=&\Delta^2\sum_{i=1}^L(\sqrt{\hat{n}_i}\:\hat{a}_i^{\dg}\hat{a}_{i+1}^{\dg}\sqrt{\hat{n}_{i+1}+1}\nn\\&+&\sqrt{\hat{n}_{i+1}+1}\:\hat{a}_{i+1}\hat{a}_{i}\sqrt{\hat{n}_{i}}).\label{HadD}
\eea
Next we calculate $\hat{H}_1^2\bullet \one$ term in Eq.~\eqref{Trotter}. Expanding $\hat{H}_1^2\bullet \one=(\hat{H}_{J}^2+\hat{H}_{J}\hat{H}_{\Delta}+\hat{H}_{\Delta}\hat{H}_{J}+\hat{H}_{\Delta}^2)\bullet \one$, we notice that $\hat{H}_{J}\hat{H}_{\Delta}$ or $\hat{H}_{\Delta}\hat{H}_{J}$ does not have diagonal elements since $\hat{H}_{J}\hat{H}_{\Delta}$, $\hat{H}_{\Delta}\hat{H}_{J}$ have either three creation operators and one annihilation operator or one creation operator and three annihilation operators. Therefore, the Hadamard product  of $\hat{H}_{J}\hat{H}_{\Delta}$ or $\hat{H}_{\Delta}\hat{H}_{J}$ with identity operator is zero. Thus,
\bea
\hat{H}_1^2\bullet \one=\hat{H}_{J}^2\bullet \one+\hat{H}_{\Delta}^2\bullet \one.\label{Had2}
\eea
The diagonal entries in $\hat{H}_{J}^2\bullet \one$ appear from the reversal of hopping terms, e.g., from combination of $\hat{a}_i^{\dg}\hat{a}_{i+1}$ and $\hat{a}_{i+1}^{\dg}\hat{a}_{i}$. After some algebra one obtains:
\bea
\hat{H}_{J}^2\bullet \one&=&2NJ^2+2J^2\sum_{i=1}^L\hat{n}_i\hat{n}_{i+1},\label{Had3}\\
\hat{H}_{\Delta}^2\bullet \one&=&2N\Delta^2+\Delta^2L+2\Delta^2\sum_{i=1}^L\hat{n}_i\hat{n}_{i+1}.\label{Had4}
\eea
We apply Eqs.~(\ref{Had1}-\ref{Had4}) to Eq.~\eqref{Trotter}, and perform some algebraic simplification to get a compact form of the following generating Hamiltonian in the continuous-time/Trotter regime, i.e., at small $J, \Delta$:
\bea
\mathcal{M}&=&(\one-\Delta^2 L-2N\mathcal{U})+\sum_{i=1}^L\Big(-2\:\mathcal{U}\hat{n}_i\hat{n}_{i+1}\nn\\&+&J^2\big(\sqrt{\hat{n}_i}\:\hat{a}_i^{\dg}\hat{a}_{i+1}\sqrt{\hat{n}_{i+1}}+\sqrt{\hat{n}_{i+1}}\:\hat{a}_{i+1}^{\dg}\hat{a}_{i}\sqrt{\hat{n}_i}\big)\nn\\&+&\Delta^2\big(\sqrt{\hat{n}_i}\:\hat{a}_i^{\dg}\hat{a}_{i+1}^{\dg}\sqrt{\hat{n}_{i+1}+1}\nn\\&+&\sqrt{\hat{n}_{i+1}+1}\:\hat{a}_{i+1}\hat{a}_{i}\sqrt{\hat{n}_i}\big)\Big)+ {\cal O}(J^4, \Delta^4),\label{MHamB}
\eea
where $\mathcal{U}=J^2+\Delta^2$. To best of our knowledge, (\ref{MHamB}) is not a well known Hamiltonian in contrast to the spin-1/2 Heisenberg model for the fermionic chain. The spectral properties of this Hamiltonian are also not known, and these are not easy to derive analytically.

\section{System-size scaling of Thouless time}\label{scal}
To uncover non-abelian symmetry of the Hamiltonian~\eqref{MHamB}, we define a set of local operators:
\bea
\hat{K}^0_i=-(\hat{n}_i+1/2),~ \hat{K}^+_i=\hat{a}_i\sqrt{\hat{n}_i},~ \hat{K}^-_i=\sqrt{\hat{n}_i}\hat{a}^{\dagger}_i,\quad
\eea
which satisfy the commutation relations of $SU(1,1)$ algebra at the same site, and commute otherwise:
\bea
[\hat{K}^+_i,\hat{K}^-_j]=-2\hat{K}^0_i \delta_{ij},~ [\hat{K}^0_i,\hat{K}^{\pm}_j]=\pm\hat{K}^{\pm}_i \delta_{ij}.  \label{eq:SU11}
\eea
The generating Hamiltonian~\eqref{MHamB} can be expressed in terms of the above operators when $\Delta=0$:
\bea
\mathcal{M}&=&\one+\sum_{i=1}^L\Big(J^2(\hat{K}^-_i\hat{K}^+_{i+1}+\hat{K}^-_{i+1}\hat{K}^+_{i})\nn\\&-&2J^2(\hat{K}^0_i\hat{K}^0_{i+1}-\f{1}{4})\Big)+{\cal O}(J^4).
\eea
The above form of $\mathcal{M}$ can be used to show 
\begin{equation}
[\hat{K}^\alpha,\mathcal{M}]=0,
\end{equation}
where $\hat{K}^\alpha = \sum_{i=1}^L \hat{K}^\alpha_i$, $\alpha\in\{+,-,0\}$ again satisfy $SU(1,1)$ algebra (\ref{eq:SU11}).
This fact indicates that the generating Hamiltonian of the Markov matrix $\mathcal{M}$ has a non-abelian $SU(1,1)$ symmetry in the particle-number conserving case of our bosonic model \cite{Giardina09,Frassek20}. We further observe by numerical checks that, when $\Delta=0$, $\mathcal{M}$ has $SU(1,1)$ symmetry for arbitrary values of $J$ beyond the Trotter regime. Such checks are  again carried out by numerically constructing the operators $K^{\alpha}$ in the Fock basis states and explicitly checking commutation of $K^{\alpha}$ with $\mathcal{M}$ for arbitrary $J$ when $\Delta=0$.

The Lie group $SU(1,1)$ is non-compact and all its unitary irreducible representations are infinite-dimensional. Due to the $SU(1,1)$ symmetry of the generating Hamiltonian, its lowest excited states can be obtained as degenerate descendants of the single-particle $(N=1)$ states, i.e., by applying the operator $\hat{K}^-$. Therefore, the $L$-dependence of $\lambda_1$ is independent of $N$ when $\Delta=0$ (see Appendix~\ref{appC} for more information). Thus we consider the case of a single boson ($N=1$) for which  (\ref{MHamB}) becomes a free boson Hamiltonian: 
\bea
\mathcal{M}|_{\Delta=0}^{N=1}&=&(\one-2J^2)+\sum_{i=1}^LJ^2(\hat{a}_i^{\dg}\hat{a}_{i+1}+\hat{a}_{i+1}^{\dg}\hat{a}_{i})+{\cal O}(J^4).\nn\\\label{MHamB1}
\eea
The ``ground state'' of the generating Hamiltonian (\ref{MHamB1}) is a state with eigenvalue 1  and with zero momentum. The eigenenergy spectrum of the Hamiltonian (\ref{MHamB1}) is gapless, and the first ``excited state'' (with momentum $k=2\pi/L$) nearest to the eigenvalue 1 goes as 
$\lambda_1=1-c_2/L^2$ where $c_2=4\pi^2 J^2$ is a constant. Thus, we find that the Thouless time depends quadratically on the length of the bosonic lattice, $t^*\simeq L^2/c_2$, for a single boson and, due to $SU(1,1)$ symmetry, for any number of bosons in the particle-number conserving model. We have numerically computed $L$-dependence of the first excited state of the Hamiltonian~\eqref{MHamB} with $N=2$ corroborating the predicted system-size scaling. The generating Hamiltonian~\eqref{MHamB} lacks $SU(1,1)$ symmetry when $\Delta \ne 0$. Consequently, the second largest eigenvalue $\lambda_1$ changes with $N$ or $N_{\rm max}$ for a fixed $L$.

We next numerically check the $L$-dependence of $\lambda_1$ of $\MM$ for arbitrary $J$ and $\Delta$. From Tab.~\ref{table1}, we find at $J=1, \Delta=0$: $\lambda_1  \sim 1-8.29/L^{1.94}$ (or $\lambda_1  \sim e^{-11.4/L^{2.05}})$ for $N/L=1/2$ (using the largest three available system sizes $L=10,12,14$), and $\lambda_1  \sim 1-9.0/L^{1.97}$ (or $\lambda_1  \sim e^{-10.5/L^{2.02}})$ for $N/L=1/4$ (using the largest three system sizes $L=12,16,20$). The above exponents for two different finite size fittings of $\lambda_1$ show a clear trend towards ${\cal O}(L^2)$ scaling of $t^*$  in the bosonic chain when $\Delta=0$. Further, the system-size scaling is also independent of number of bosons in the chain in the presence of $U(1)$ symmetry for arbitrary $J$ as predicted above due to $SU(1,1)$ symmetry of $\MM$. The last observation is clear from the fact that the value of $\lambda_1$ in Tab.~\ref{table1} is the same for two different $N/L$ at any particular $L$.  We have also numerically computed raw SFF $K(t)$ using definition~\eqref{SFF} for different $L$, which confirms our analytical prediction based on RPA for the $L$-dependence of $t^*$ when $\Delta=0$ (see Appendix~\ref{appA} for details).

\begin{table}[h!]
  \begin{center}
    \begin{tabular}{c|c|c|c|c|c|c|c}
      \multicolumn{4}{c|}{$J=1,\Delta=0, N/L=1/2$} & \multicolumn{4}{c}{$J=1,\Delta=0, N/L=1/4$}\\ 
      \hline
      $L$ & $\lambda_1$ & $\lambda_2$ & $\lambda_3$ & $L$ & $\lambda_1$ & $\lambda_2$ & $\lambda_3$\\
      \hline
      8  & 0.8526 & 0.7486 & 0.6680 & 8 & 0.8526   & 0.7486 & 0.4847\\ 
      10 & 0.9042 & 0.8283 & 0.7658 &12 & 0.9329   & 0.8764 & 0.8278\\
      12 & 0.9329 & 0.8764 & 0.8278 &16 & 0.9619   & 0.9278 & 0.8970\\
      14 & 0.9504 & 0.9071 & 0.8688 &20 & 0.9755  & 0.9529 & 0.9320\\
    \end{tabular}
     \caption{Three largest eigenvalues $\lambda_1,\lambda_2,\lambda_3$ (excluding $\lambda_0=1$) of $\mathcal{M}$ for various lengths $L$ at two different filling fractions $N/L$ in the presence of $U(1)$ symmetry in the kicked interacting bosonic chain. The value of $\lambda_1$ is independent of $N/L$ for any particular $L$. }
    \label{table1}
  \end{center}
\end{table}

\begin{figure}
\includegraphics[width=\linewidth]{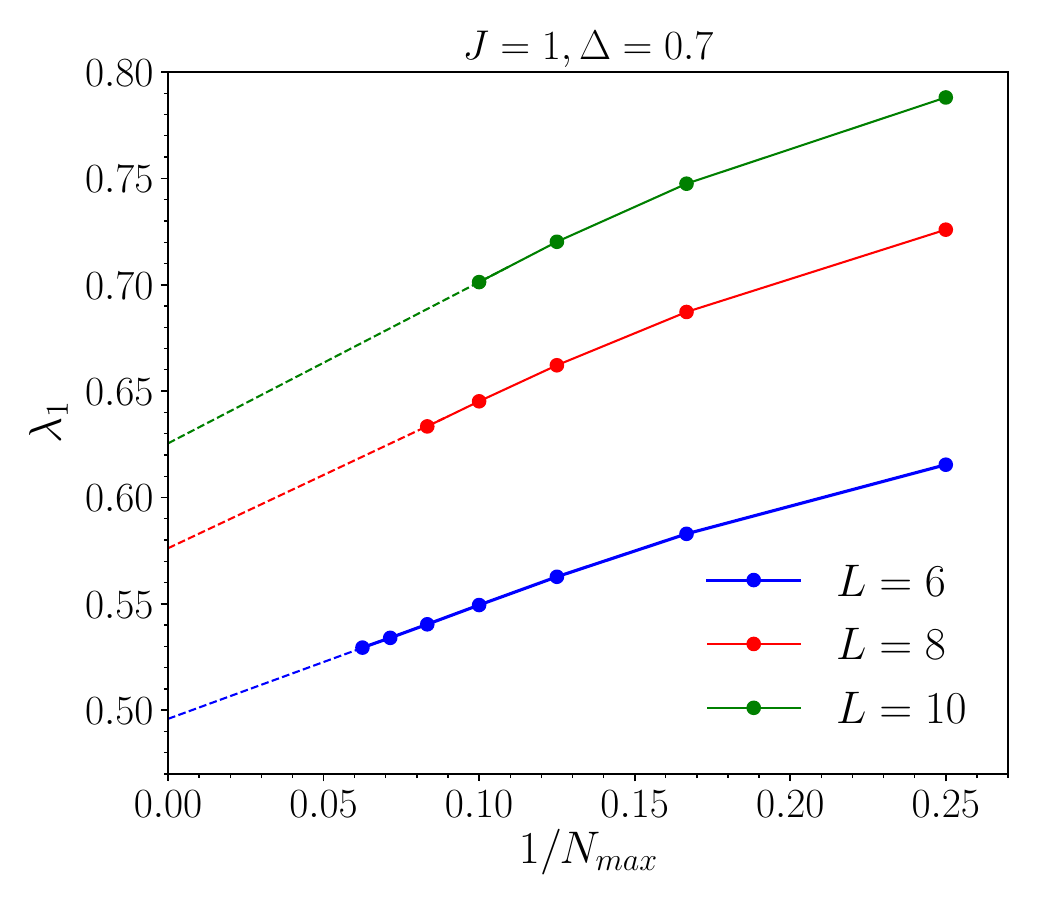}
\caption{Second largest eigenvalue $\lambda_1$ of $\mathcal{M}$ with the inverse of maximum number of bosons $(1/N_{\rm max})$ for three lengths $(L)$ in the absence of $U(1)$ symmetry in the kicked interacting bosonic chain. The dashed lines indicate a linear extrapolation of the last few large $N_{\rm max}$ points. These linear extrapolations give  $\lambda_1 \sim 1-1.43/L^{0.58}$ or $e^{-2.89/L^{0.79}}$ at $1/N_{\rm max} \to 0$, which indicates a finite system-size dependence of the Thouless time for the bosonic chain when $\Delta \ne 0$.}
\label{BosNonConI}
\end{figure}

On the other hand, it is very challenging to find the $L$-dependence of $\lambda_1$ in the bosonic chain when $\Delta \ne 0$ as $\mathcal{N}$ is formally infinite for any $L$. Nevertheless, we vary truncation number $N_{\rm max}$ and $\mathcal{N}$ for a fixed $L$ to get an estimate of $\lambda_1$ in the large $N_{\rm max}$ limit.
Using clear linear extrapolations in $1/N_{\rm max}$ towards $1/N_{\rm max}=0$, shown in Fig.~\ref{BosNonConI}, we find strong evidence for a nontrivial $L$-dependence of $\lambda_1$, and $t^* = {\mathcal O}(L^\gamma)$, $\gamma=0.7\pm 0.1$, in the bosonic chain for $J=1,\Delta = 0.7$. 
The last result markedly differs from the ${\cal O}(L^0)$ scaling of $t^*$ in the absence of $U(1)$ symmetry in the fermionic chain. We further observe from our numerics with limited system sizes that the $L$-dependence of $\lambda_1$ seems to be close to the above $\gamma$ value even when $\Delta$ is tuned a bit, which we show in Appendix~\ref{appB}.  

\section{Summary and outlook}\label{sum}
In summary, we reported on discovering universal non-abelian symmetries of the Markov matrices whose subleading eigenvalues determine the system-size scaling of Thouless time to reach universal RMT form for SFF in correlated bosonic and fermionic chains with periodic driving (kicking). These symmetries lead to identical quantum chaotic features in the studied bosonic and fermionic models in the presence of particle-number conservation. Without particle-number conservation, the fermionic and bosonic models display different system-size scaling of the Thouless time. The proposed bosonic model is convenient for experimental realization with photons in various engineered optical systems \cite{RoyRMP2017, scully1997quantum, EckardtRMP2016}, which can be applied to verify our predictions. Both for fermions and bosons, our estimate for the system-size dependence of Thouless time using the second-largest eigenvalue of $\mathcal{M}$ derived within the RPA shows a good agreement with that from the directly simulated $K(t)$ using the definition in Eq.~\ref{SFF} for different $L$ and $\Delta$. Such agreement is achieved since the $L$-dependence of the third-largest eigenvalue $\lambda_2$ (along with the successive few largest eigenvalues) is the same as the second-largest eigenvalue $\lambda_1$ both for fermions and bosons when longer lengths and finite filling fractions are considered. Thus, the qualitative features of the universal form of $K(t)$ and $t^*$ are not affected due to the restriction of the analysis to the second largest eigenvalue (see Appendix~\ref{appA} for discussion on the nonuniversal part of $K(t)$).

Nevertheless, it is necessary and exciting to find second- or higher-order contributions to the leading order RMT form of $K(t)=2t$ derived in this paper. A second-order of $t/t_H$ term for the universal RMT form of the COE was calculated for periodically kicked transverse-field Ising model in \cite{KosPRX2018} by going beyond the identity permutation in writing Eq.~\ref{SP}. Such a derivation would be more challenging for our generic model of fermions or bosons studied in Ref.~\cite{RoyPRE2020} and the current manuscript. We hope to pursue such calculation for the current model in future studies.  
 
\section{Acknowledgment}
DR thank R. Singh for discussions. DR acknowledges funding from the Ministry of Electronics $\&$ Information Technology (MeitY), India under the grant for ``Centre for Excellence in Quantum Technologies" with Ref. No. 4(7)/2020-ITEA.
TP acknowledges support by European Research Council (ERC) under Advanced grant 694544-OMNES, and by Slovenian Research Agency (ARRS) under program P1-0402.

\begin{figure}[h!]
\includegraphics[width=\linewidth]{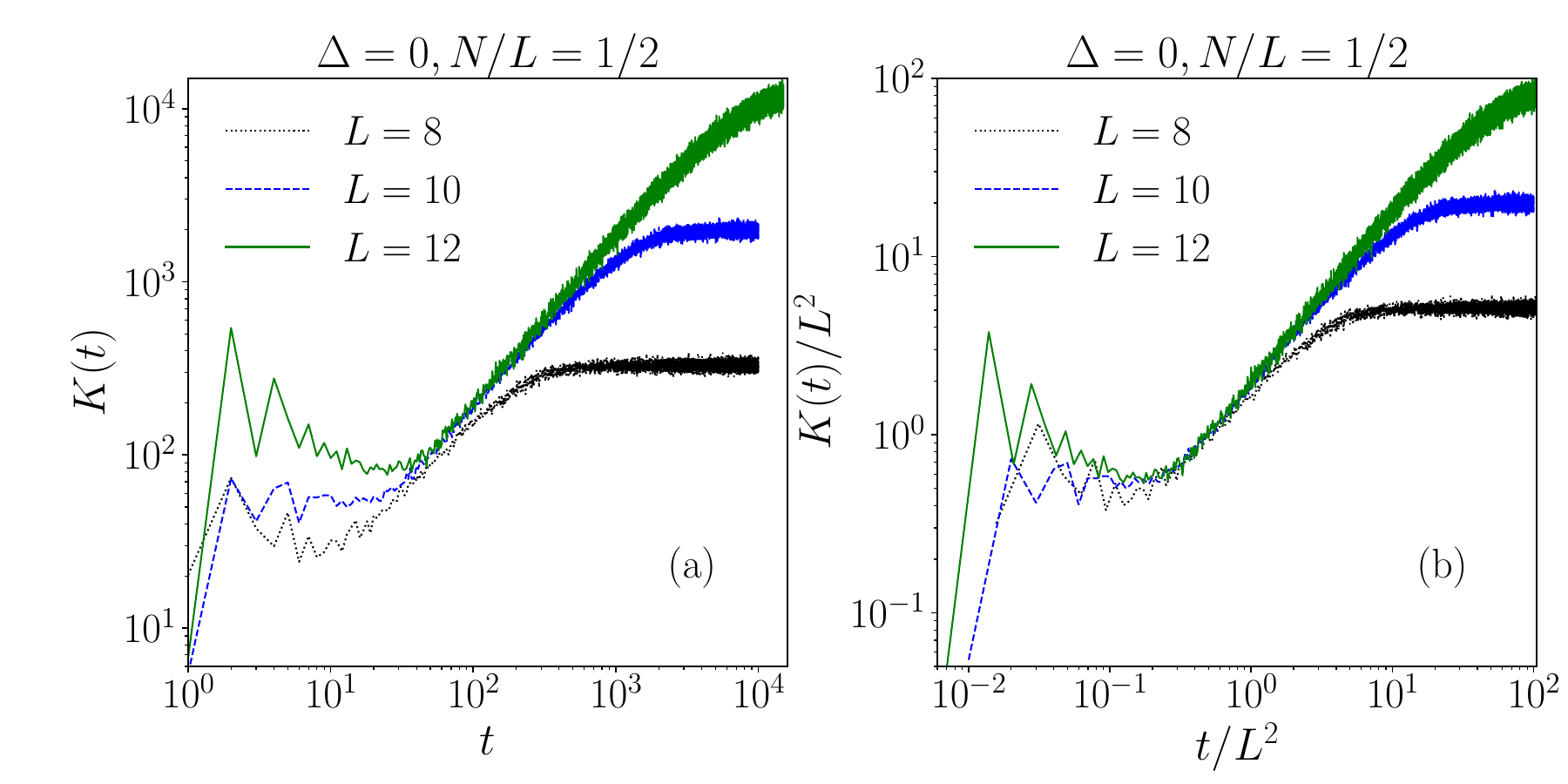}
\caption{Spectral form factor $K(t)$ at half filling for different system sizes $L$ of the kicked spinless boson chain with particle-number conservation. Here, $J=1, U_0 = 15, \alpha = 1.5, \Delta=0$. We use the open boundary conditions in real space, and take $N/L = 1/2$. An averaging over 200 to 500 realizations of disorder is performed. In (b), we show the data collapse in scaled time $t/L^2$. A longer $L$ has a higher saturation value of $K(t)$.}
\label{SFFHF}
\end{figure}
\appendix
\section{Exact numerical computations of $K(t)$ using Eq.~\ref{SFF} for $\Delta=0$}\label{appA}
\begin{figure}[h!]
\includegraphics[width=\linewidth]{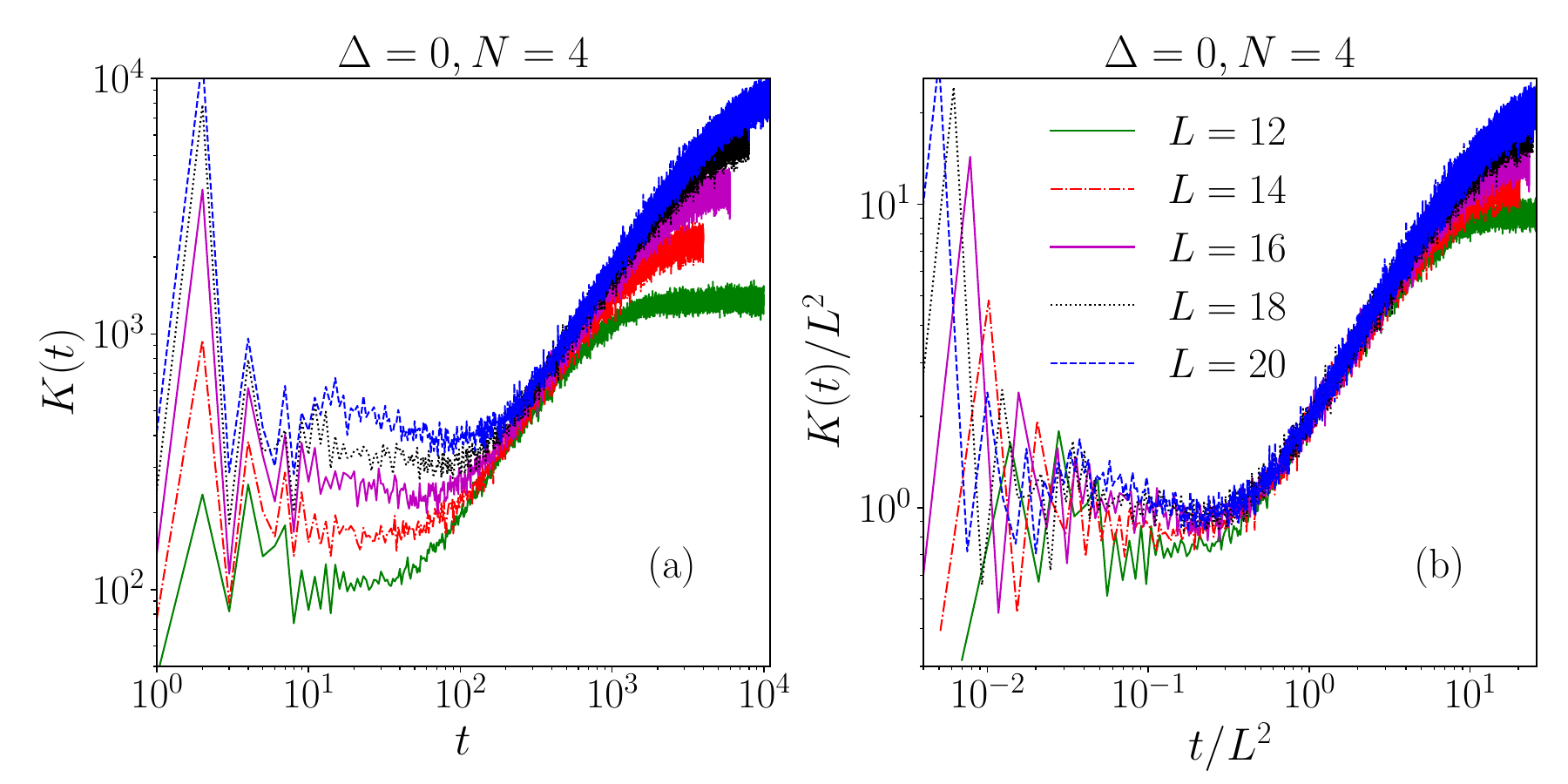}
\caption{Spectral form factor $K(t)$ at fixed number of bosons for different system sizes $L$ of the kicked spinless boson chain with particle-number conservation. Here, $J=1, U_0 = 15, \alpha = 1.5, \Delta=0$. We use the open boundary conditions in real space, and take $N = 4$ for all $L$. An averaging over 200 to 500 realizations of disorder is performed. In (b), we show the data collapse in scaled time $t/L^2$.}
\label{SFFN4}
\end{figure}
In Eq.~\ref{RMT} and Sec.~\ref{scal}, we have predicted the system-size scaling of $t^*$ in the $U(1)$ symmetric kicked bosonic chain using the Hamiltonian form of $\MM$ and the numerical finding of $\lambda_1$ of $\MM$ for different $L$. To further validate this prediction by applying the RPA and the identity permutation approximation, we numerically calculate $K(t)$ directly using Eq.~\ref{SFF} for our model in Eq.~\ref{ham1}. We show the numerically computed $K(t)$ for $\Delta=0$ in Fig.~\ref{SFFHF} for half filling and in Fig.~\ref{SFFN4} for a fixed number of bosons in the chain. The initial temporal growth of $K(t)$ at $t \ll t_H$ in Figs.~\ref{SFFHF}(a) and \ref{SFFN4}(a) depends strongly on $L$, and $K(t)$ further grows linearly with time before saturating around $t_H=\mathcal{N}$, which depends on the number of bosons $N$ in the chain. We plot $K(t)/L^2$ with $t/L^2$ in Figs.~\ref{SFFHF}(b) and \ref{SFFN4}(b) to find the $L$-dependence of the initial temporal growth of the SFF. We  obtain a good data collapse for various $L$ and $N/L$, which shows an agreement with the predicted $L$-dependence of $K(t)$ for the particle-number conserving bosonic chain. Therefore, we confirm that the analytical predictions using the RPA agree with the bosonic chain's direct numerical analysis in the presence of $U(1)$ symmetry. For the numerics, we choose a long-range form of the interaction (e.g., $\alpha=1.5$), which ensures the applicability of the RPA for approximating the phases $\theta_{\underline{n}}$ of different many-particle basis $|\underline{n}\rangle$ as independent and uniformly distributed random numbers. The values of $\alpha$ for a nondegenerate spectrum of $\hat{H}_0$ and consequently the validity of RPA have been carefully investigated in Ref.~\cite{KosPRX2018} by numerically calculating  $K(t)$ for different $\alpha$'s.

\begin{figure}[h!]
\includegraphics[width=\linewidth]{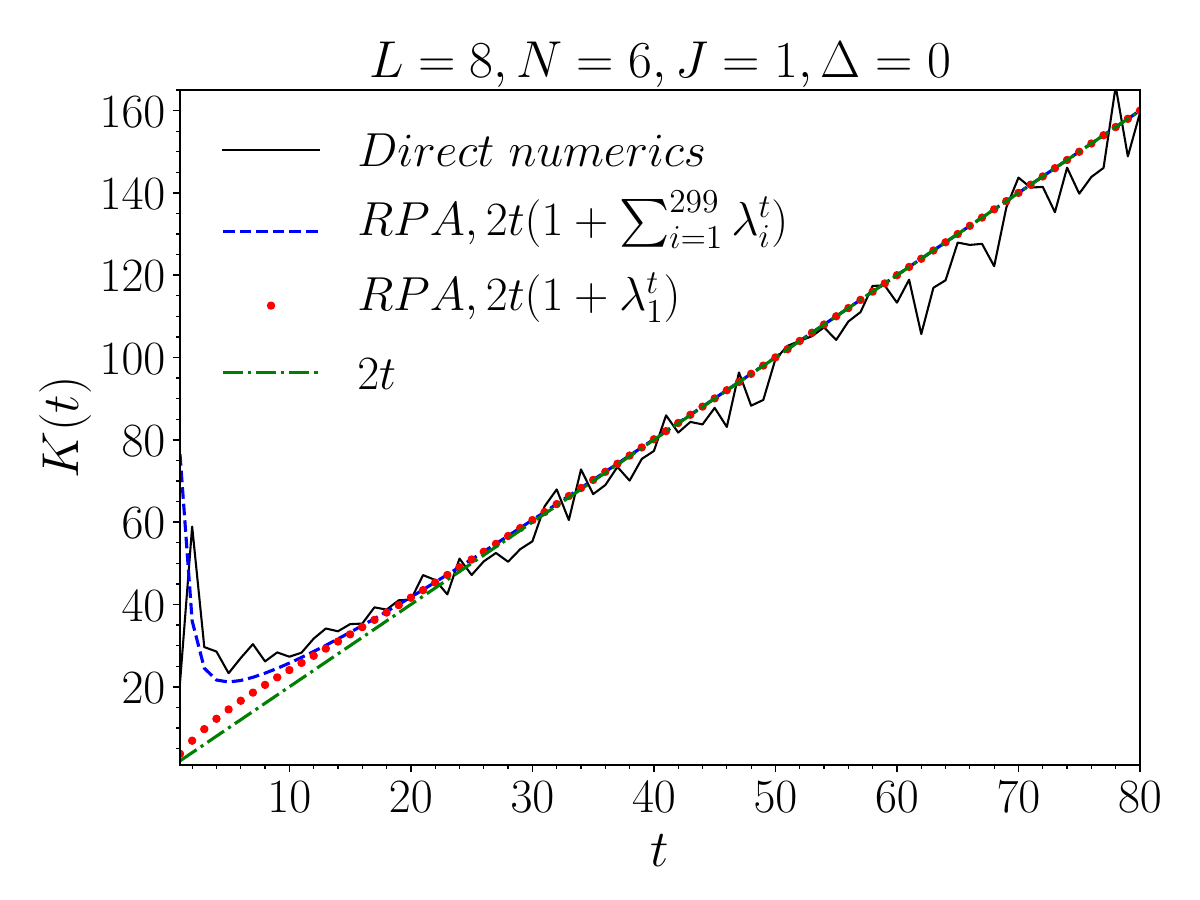}
\caption{Comparison of directly computed $K(t)$ (black line) with that obtained employing the RPA and the identity permutation approximation. $K(t)$ within the RPA is shown by keeping up to second largest eigenvalue (red dots) and by keeping $300$ largest eigenvalues (blue dashes). While blue dashes capture both the nonuniversal and universal parts of $K(t)$, the red dots only match with the universal RMT form of $K(t)$ beyond the Thouless time. The direct numerical computation of $K(t)$ is carried out with $U_0 = 15, \alpha = 1.5$.} 
\label{NonUniv}
\end{figure}
In Fig.~\ref{NonUniv}, we compare the directly computed $K(t)$ with that obtained by employing the RPA and the identity permutation approximation. We show $K(t)$ within the RPA by keeping up to the second largest eigenvalue and by keeping $300$ largest eigenvalues. The SFF calculated within the RPA can capture both the nonuniversal part of $K(t)$ for a short time and the universal part of $K(t)$ beyond $t^*$ when a significant fraction of eigenvalues $\lambda_i$ of $\mathcal{M}$ (e.g., largest $300$ out of total $1716$ of $\lambda_i$ for parameters in Fig.~\ref{NonUniv}) is included in Eq.~\ref{SP}. The form $K(t)$ in Eq.~\ref{RMT} by keeping up to the second largest eigenvalue $\lambda_1$ matches with the universal RMT form of $K(t)$ beyond the Thouless time as shown in Fig.~\ref{NonUniv}.

\section{System-size scaling of $\lambda_1$ of $\mathcal{M}$ for $\Delta \ne 0$}
\label{appB}
\begin{figure}[h!]
\includegraphics[width=\linewidth]{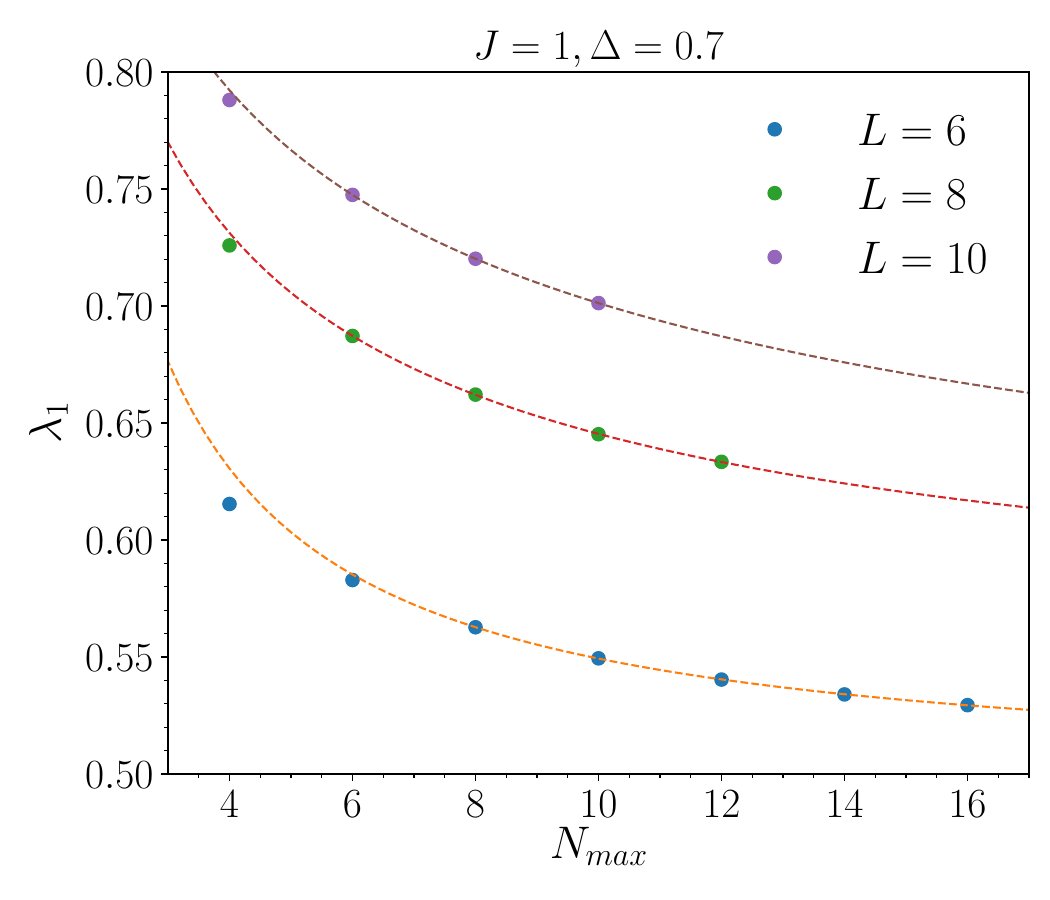}
\caption{Second largest eigenvalue $\lambda_1$ of $\mathcal{M}$ with increasing maximum  number of bosons $N_{\rm max}$ for three lengths $L$ in the absence of $U(1)$ symmetry in the kicked interacting bosonic chain. The dashed lines indicate an asymptotic behavior of the limited data for longer $L=8,10$. These data points along with asymptotic lines show $\lambda_1 \sim 1-1.6/L^{0.68}$ or $e^{-3.13/L^{0.89}}$ at $N_{\rm max}=16$, which predicts a finite system-size dependence of the Thouless time for the bosonic chain when $\Delta \ne 0$.}
\label{BosNonCon}
\end{figure}

We here discuss the $L$-dependence of the second largest eigenvalue $\lambda_1$ of $\mathcal{M}$ for various pairing strength $\Delta$ in the absence of $U(1)$ symmetry in the bosonic model. In Fig.~\ref{BosNonConI}, we have shown $\lambda_1$  with $1/N_{\rm max}$ for $L=6,8,10$, and $J=1,\Delta=0.7$. We could calculate an estimate for the asymptotic feature of $\lambda_1$ at large $N_{\rm max}$ (or small $1/N_{\rm max}$) by linearly extrapolating the last few points. Such an estimate gives finite system-size scaling of $\lambda_1$ and $t^*$ with $L$ for large values of $J$ and $\Delta$. We further display $\lambda_1$ with increasing $N_{\rm max}$ (instead of $1/N_{\rm max}$ in Fig.~\ref{BosNonConI}) in Fig.~\ref{BosNonCon} for the same parameters as in Fig.~\ref{BosNonConI}. The dashed lines in Fig.~\ref{BosNonCon} give an asymptotic behavior of the limited data for longer $L=8,10$. We find $\lambda_1 \sim 1-1.6/L^{0.68}$ or $e^{-3.13/L^{0.89}}$ at $N_{\rm max}=16$, using these data points and asymptotic lines. 

\begin{figure}[h!]
\includegraphics[width=\linewidth]{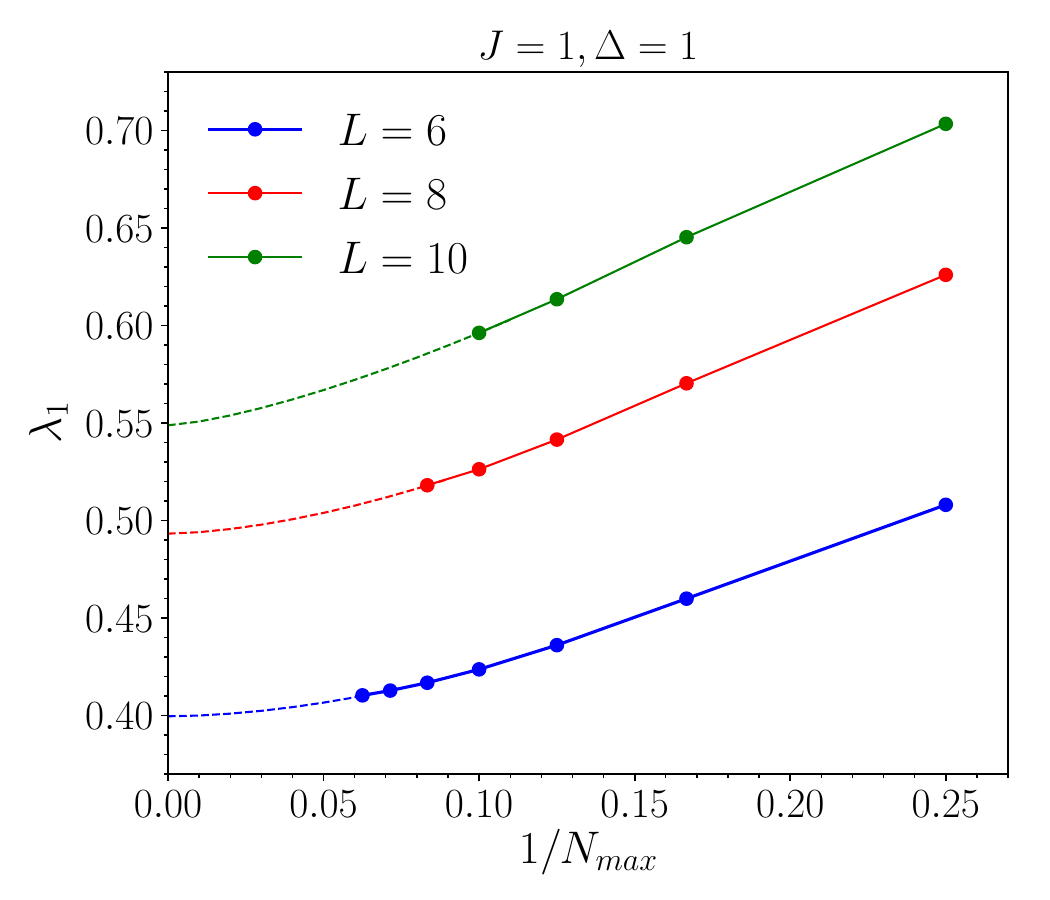}
\caption{Second largest eigenvalue $\lambda_1$ of $\mathcal{M}$ with the inverse of maximum number of bosons, $1/N_{\rm max}$, for three lengths $L$ at $J=\Delta=1$ in the absence of $U(1)$ symmetry in the kicked interacting bosonic chain.  The dashed lines denote algebraic extrapolation of the last few large $N_{\rm max}$ points. These extrapolations give $\lambda_1 \sim 1-1.65/L^{0.56}$ or $e^{-4.09/L^{0.83}}$ at $1/N_{\rm max} \to 0$.}
\label{BosNonConD}
\end{figure}

In Fig.~\ref{BosNonConD}, we show the $L$-dependence of $\lambda_1$ for a larger $\Delta~(=1)$, which gives  $\lambda_1 \sim 1-1.65/L^{0.56}$ or $e^{-4.09/L^{0.83}}$ at $1/N_{\rm max} \to 0$ by using the extrapolations for large $N_{\rm max}$. Thus, we have $\gamma=0.7\pm 0.13$ when $J=\Delta=1$. Therefore, the $L$-dependence of $\lambda_1$ and $t^*$ mostly remains the same for different values of finite $\Delta$. This $L$-dependence of $\lambda_1$ in the bosonic chain is clearly different from the $L$-independence of $\lambda_1$ in the fermionic model in the absence of $U(1)$ symmetry.

\section{Nonabelian symmetries of $\mathcal{M}$ for $\Delta=0$}
\label{appC}
\begin{figure}[h!]
\includegraphics[width=\linewidth]{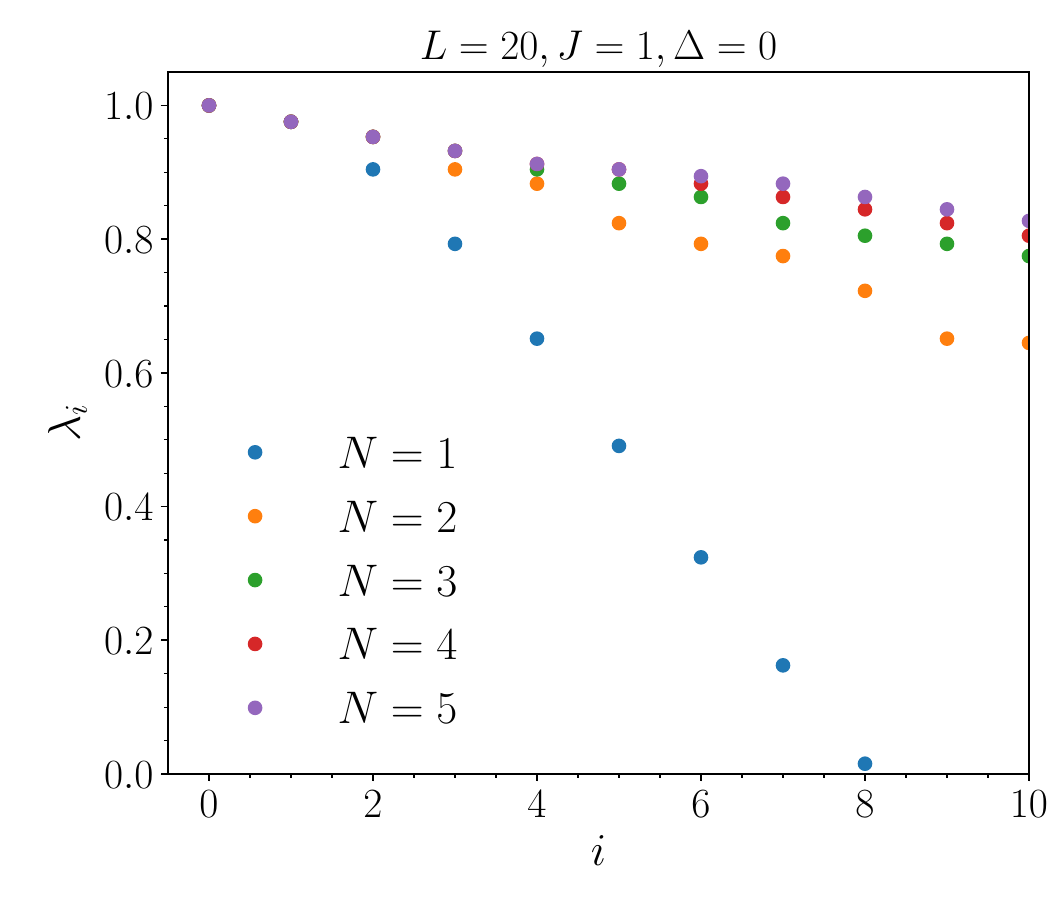}
\caption{Largest eigenvalues $\lambda_i,~i \in \{0,1,2,\dots,9\}$ of $\mathcal{M}$ for different number $N$ of bosons at $J=1, \Delta=0$ of the kicked interacting bosonic chain of length $L=20$.  The largest two eigenvalues are the same for all non-zero $N$, and at least $m$ largest eigenvalues are the same for all $N \ge (m-1)$ due to $SU(1,1)$ symmetry of $\mathcal{M}$.}
\label{SpeUniSymm}
\end{figure}

When the total particle number is conserved (e.g., $\Delta=0$) in our fermionic and bosonic model, we have found, respectively, $SU(2)$ and $SU(1,1)$ symmetry of the Markov matrix $\mathcal{M}$ for arbitrary $J$ and the Hamiltonian form of $\mathcal{M}$ at small $J$. The presence of such symmetries for the Hamiltonian form of $\mathcal{M}$ implies that the ``lowest excited states'' of the Hamiltonian are related for the different number of fermions or bosons in the model, i.e., they represent degenerate symmetry multiplets. In Fig.~\ref{SpeUniSymm}, we explicitly compare the ten largest eigenvalues of  $\mathcal{M}$ (which are related to ``lowest excited states'' of the Hamiltonian form of $\mathcal{M}$) with different $N$'s, including $N=1$ for a fixed length of the bosonic chain. We find from Fig.~\ref{SpeUniSymm} that while the largest two eigenvalues $(\lambda_0,\lambda_1)$ are the same for all $N$ including $N=1$, at least $m$ largest eigenvalues $(\lambda_i,~i \in \{0,1,2,\dots,m-1\})$ are the same for all $N \ge (m-1)$ due to $SU(1,1)$ symmetry of $\mathcal{M}$.

\bibliography{bibliographyRMT}

\end{document}